\documentclass[12pt]{iopart}
\usepackage{iopams}
\usepackage{graphicx}
\usepackage{amssymb}

\usepackage{color}

\begin{document}

\title[Classical to quantum crossover of the cyclotron resonance in graphene]
{Classical to quantum crossover of the cyclotron resonance in graphene:\\ A study of the strength of intraband absorption}

\author{M~Orlita$^{1,2,}$\footnote[6]{Author to whom any correspondence should be addressed.}, I~Crassee$^{3}$, C~Faugeras$^1$, A~B~Kuzmenko$^3$, F~Fromm$^4$, M~Ostler$^4$,
Th~Seyller$^4$, G~Martinez$^1$, M~Polini$^5$ and M~Potemski$^1$}

\address{$^1$~Laboratoire National des Champs Magn\'etiques Intenses, CNRS-UJF-UPS-INSA, 25, avenue des Martyrs, 38042 Grenoble, France}
\address{$^2$~Faculty of Mathematics and Physics, Charles University, Ke Karlovu 3, 121 16 Prague 2, Czech Republic}
\address{$^3$~D\'{e}partement de Physique de la Mati\`{e}re Condens\'{e}e, Universit\'{e} de Gen\`{e}ve, CH-1211 Gen\`{e}ve 4, Switzerland}
\address{$^4$~Lehrstuhl f\"{u}r Technische Physik, Universit\"{a}t Erlangen-N\"{u}rnberg, D-91058 Erlangen, Germany}
\address{$^5$~NEST, Istituto Nanoscienze-CNR and Scuola Normale Superiore, I-56126 Pisa, Italy}
\ead{milan.orlita@lncmi.cnrs.fr}

\begin{abstract}
We report on absolute magneto-transmission experiments on highly-doped quasi-free-standing epitaxial graphene targeting
the classical-to-quantum crossover of the cyclotron resonance. This study allows us to directly extract the carrier density and also other relevant quantities such as the quasiparticle velocity and the Drude weight, which is precisely measured from the strength of the cyclotron resonance.
We find that the Drude weight is renormalized with respect to its non-interacting (or random-phase-approximation) value and that the renormalization is tied to the quasiparticle velocity enhancement. This finding is in agreement with recent theoretical predictions, which attribute the renormalization of the Drude weight in graphene to the interplay between broken Galilean invariance and electron-electron interactions.    
\end{abstract}


\maketitle

\section{Introduction}

The low-frequency dynamical conductivity, $\sigma(\omega)$, a
fundamental property of electrical conductors, is also central for
the physics of graphene and for possible applications of this
material in opto-electronics~\cite{CastroNetoRMP09,PeresRMP10,BonaccorsoNaturePhotonics10,DasSarmaRMP11}.
A tempting description of
$\sigma(\omega)$ in graphene refers to a simple model of noninteracting Dirac-fermion quasiparticles with the characteristic
linear dispersion $E({\mathbf{k}})=\pm v_{\rm F}\hbar|{\mathbf{k}}|$, which at zero temperature occupy the
electronic states up to the Fermi energy $|E_{\rm F}|= \hbar
v_{\rm F}\sqrt{\pi |n|}$, where $v_{\rm F}$ stands for the (Fermi)
velocity parameter and $n$ is the free-carrier concentration. Such
an effective single-particle (ESP) picture correctly reproduces
the measured strength of interband-absorption processes with the
characteristic onset at $2|E_{\rm F}|$ and the dispersionless
universal amplitude $\sigma_{\rm inter}=\sigma_{\rm uni} = e^2/(4\hbar)$~\cite{NairScience08,MakPRL08,WangScience08,KuzmenkoPRL08,StauberPRB08,FeiPRB08,LiNaturePhys08}.
Further simple reasoning in terms of Drude approach yields the
well-known expression for the integrated strength ${\cal
D}=\int_{0} ^{\infty } \sigma_{\mathrm{intra}} (\omega) d\omega$ of
intraband processes (so-called Drude weight):
\begin{equation}\label{eq:singleparticlelike}
{\cal  D} =  \frac{e^2}{2\hbar} v_{\rm F} \sqrt{\pi |n|}=\frac{\pi |n|
e^2}{2m_{\rm c}}=\frac{2\sigma_{\rm uni} |E_{\rm F}|}{\hbar},
\end{equation}
in which $m_{\rm c}= |E_{\rm F}|/v ^2_{\rm F}$ accounts for the
effective cyclotron mass of carriers at the Fermi energy.

The experimental measurements available so far~\cite{HorngPRB11,YanACSNano11} are
in conflict with this simple prediction. They suggest
a {\it suppression} of the Drude weight up to 40~$\%$ (notably, in graphenes with high carrier
concentrations), and therefore raise the issue of the
applicability of the ESP picture to describe the dynamical
conductivity of graphene. These doubts are additionally fostered
by a startling report on non-vanishing absorption below the
$2|E_{\rm F}|$ threshold for interband processes~\cite{LiNaturePhys08}.

Formally, the Drude approach follows from the commonly used random phase
approximation (RPA)~\cite{Giuliani_and_Vignale}, which indeed grants Eq.~(\ref{eq:singleparticlelike}), however, with the
velocity parameter fixed at its bare value~\cite{AbedinpourPRB11} $v_{{\rm F}, 0}= 0.85 \times
10^{6}~{\rm m}/{\rm s}$ -- the value of $v_{{\rm F},0}$ can be accurately estimated, for example,  from
calculations based on density-functional theory at the LDA level~\cite{YangPRL09,SchilfgaardePRB11,AttaccalitePSSB09}. Markedly,
$v_{{\rm F}, 0}$ is smaller than the apparent (renormalized) Fermi
velocity, $v_{\rm F}= (1.0 - 1.1) \times 10^{6}~{\rm m}/{\rm s}$, which is derived from
most of spectroscopy studies on graphene-based
systems~\cite{SadowskiPRL06,JiangPRL07,DeaconPRB07,OrlitaPRL08II,OrlitaSST10,HenriksenPRL10,CrasseePRB11,OrlitaPRL11,BooshehriPRB12} (and from beyond-LDA theoretical approaches).
In fact, the Drude weight determined by the bare velocity, ${\cal  D}_0 = (v_{{\rm F}, 0}/v_{\rm F}){\cal  D}$, could partially explain the experimental claims~\cite{HorngPRB11,YanACSNano11} of a suppressed ${\cal  D}$. However, the RPA approach leading to Eq.~(\ref{eq:singleparticlelike})
with $v_{\rm F}$ replaced by $v_{{\rm F}, 0}$ misses some important physics~\cite{AbedinpourPRB11}.
A more adequate formalism, based on many-body diagrammatic perturbation
theory~\cite{AbedinpourPRB11}, provides a more complex scenario in which self-energy ({\it i.e.}, velocity enhancement) and vertex ({\it i.e.}, excitonic) corrections compete with each other.
Nevertheless, when the carrier concentration is sufficiently high, the
approximate validity of the ESP model of noninteracting Dirac-fermion quasiparticles is expected to be
recovered and Eq.~(\ref{eq:singleparticlelike}), with the
renormalized (apparent) velocity parameter, may be justified. Thus
the issue of the strength of the Drude weight in graphene, one of
its very basic electronic properties, remains a matter of
controversy.

In this paper we report on testing the ESP model against the
results of cyclotron-resonance (CR) absorption experiments,
performed on highly-doped graphene specimens. Consistently following the ESP
picture, we notice that the Drude weight is fully transformed into
the strength of the CR absorption in the classical limit and, with
a good precision, into the strength of inter-Landau-level
transitions in the quantum limit, {\it i.e.}, into a quantity which is
accurately determined from our absolute magneto-transmission
measurements. The carrier concentration is deduced from the analysis
of the oscillatory broadening
of the CR response with Landau-level filling factor, whereas the effective mass (and,  in turn,
the apparent Fermi velocity) is directly read from the dispersion
of the CR with magnetic field. The Drude weight, as
measured  from the strength of the CR absorption, is found to be
consistent with the estimate given by Eq.~(\ref{eq:singleparticlelike})
in which the velocity parameter has the meaning of its apparent (spectroscopically
measured or theoretically renormalized) value.
\begin{figure}[b]
\begin{center}
\scalebox{0.9}{\includegraphics{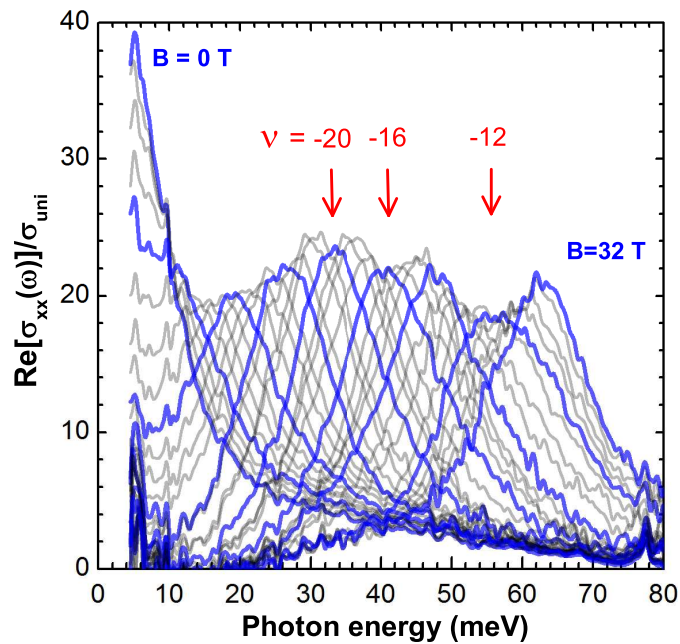}} \caption{\label{SPKT}
The real part of the longitudinal conductivity
$\sigma_{xx}(\omega,B)$ (in units of $\sigma_{\mathrm{uni}}$) as
extracted from the absolute magneto-transmission $T(B)$ using the
procedure described in Ref.~\cite{SMF}. Individual
curves are plotted every $1~{\rm T}$. Thick curves are every $4~{\rm T}$.}
\end{center}
\end{figure}

\section{Experimental results}

The investigated graphene sheet was prepared by thermal
decomposition of the Si-terminated substrate of 6H-SiC. Subsequent
hydrogenation of the ``zero layer" yields a quasi-freestanding
epitaxial graphene monolayer~\cite{RiedlPRL09}. Such specimens are
typically highly $p$-doped with the Fermi energy $E_{\rm F}\approx-0.3~{\rm eV}$ and carrier mobility $\mu \approx
(2-3)\times10^3~{\rm cm}^2/({\rm V.s})$~\cite{CrasseeNP11}. In this paper, we present data
from one particular specimen, however, consistent results have been obtained on another similar sample.

To measure the far-infrared transmittance, a macroscopic area of the sample ($\approx 4~{\rm mm}^2$) was exposed to
the radiation of a globar, which was analyzed by a Fourier transform
spectrometer and delivered to the sample via light-pipe optics. The light was detected by a
composite bolometer placed directly below the sample, kept at $T=1.8$~K. From the
substrate-normalized transmission $T(B)$, the real part of the optical
conductivity, $\Re e~[\sigma_{xx}(\omega,B)]$, was obtained for
every separate magnetic field. The extraction of the conductivity from
the transmission is described in detail in the Appendix.
\begin{figure}
\begin{center}
\scalebox{0.75}{\includegraphics*[25pt,295pt][338pt,830pt]{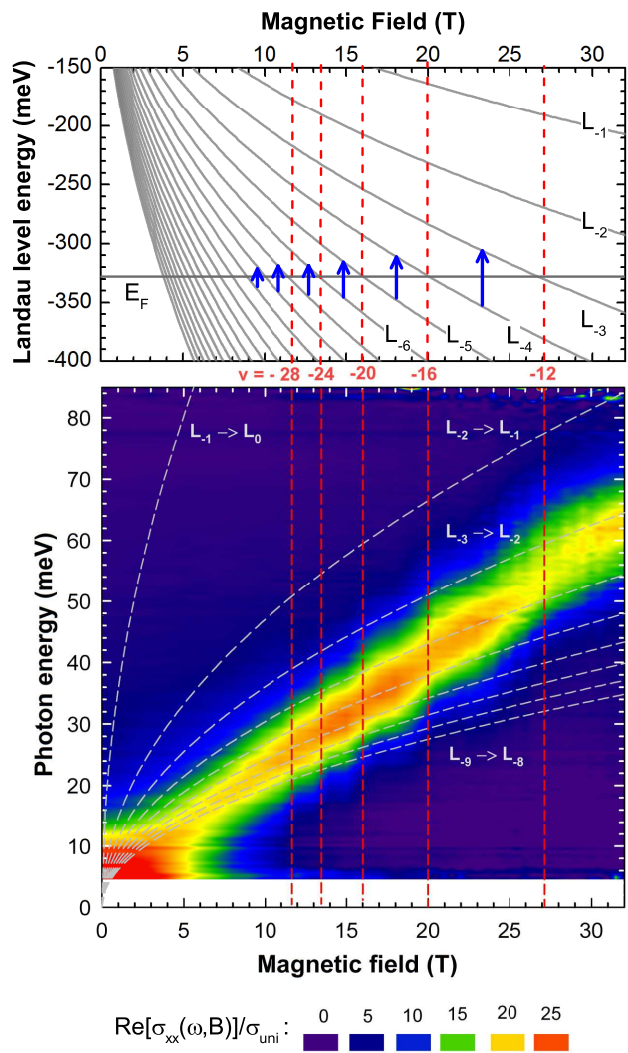}}
\caption{\label{ColorPlot}
Upper panel: Landau level fan chart with schematically-shown cyclotron resonance transitions in the quantum regime.
Lower panel: A color plot of the real part of the experimentally determined longitudinal optical conductivity $\sigma_{xx}(\omega,B)$.
The dashed lines correspond to transitions between adjacent (hole) Landau levels in graphene, $L_{-m} \rightarrow L_{-m+1}$.
 In both panels, a Fermi velocity
$v_{\rm F} = 0.99 \times 10^6~{\rm m}/{\rm s}$ was considered.}
\end{center}
\end{figure}

\begin{figure}[t]
\begin{center}
\scalebox{0.5}{\includegraphics{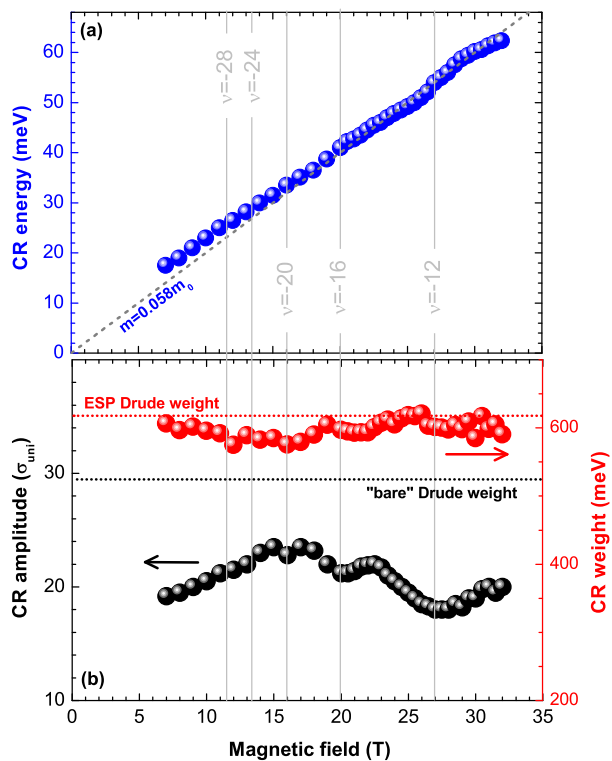}}
      \caption{\label{Fitting}
Part (a): Maximum of the CR absorption as a function of the magnetic field. The dashed line denotes the theoretical position of CR
for the effective mass of
$m=0.058\,m_0$. The departure of the CR position from the linear-in-$B$ behavior at low fields is consistent with the appearance of confined magneto-plasmons~\cite{CrasseeNL12,YanNL12}.
Part (b): Drude weight (area) extracted using a Drude-Lorentzian formula (see Refs. \cite{CrasseeNL12,YanNL12}) and the directly read-out amplitude of $\Re e~[\sigma_{xx}(\omega,B)$].
The horizontal dotted lines show the expected ESP and ``bare'' Drude weights for a specimen with a $95\%$ graphene coverage (see the main text), taking account the apparent (renormalized)
$v_{\rm F}=0.99\times 10^6~{\rm m}/{\rm s}$ and bare $v_{{\rm F}, 0} = 0.85\times 10^6~{\rm m}/{\rm s}$ Fermi velocities, respectively.}
\end{center}
\end{figure}

Experimental data are presented in Figs.~\ref{SPKT} and \ref{ColorPlot} in the form of the conductivity spectra
and as a false-color plot of $\Re e~[\sigma_{xx}(\omega,B)]$, respectively.
In agreement with previous
studies on highly-doped graphene~\cite{WitowskiPRB10,CrasseeNP11}, the magneto-optical response corresponds to the
quasi-classical regime of the CR~\cite{SchliemannNJP08,GusyninNJP09,MikhailovPRB09}, since the condition $\mu B\sim 1$ applies. The CR line is
approximately a single Lorentzian, its position in the spectra increasing linearly with $B$, see Fig.~\ref{ColorPlot}.
At low magnetic fields (in particular below $B=10$~T), a departure from linearity is visible,
see Fig.~\ref{Fitting}(a). In this regime the optical response
is dominated by plasmons confined on the scale of a few microns due to graphene grain boundaries
and/or substrate terraces. This regime has been discussed in detail elsewhere~\cite{CrasseeNL12}.

In high magnetic fields, the CR absorption still follows an overall linear-in-$B$ evolution, consistently with moderate mobility of the sample. However,
at the same time, it shows first clear indications of the approaching quantum regime, in which CR absorption
corresponds to transitions between well-defined adjacent Landau levels (LLs). A number of CR studies of graphene
in this regime has already been performed, see, {\it e.g.}, Refs.~\cite{SadowskiPRL06,JiangPRL07,DeaconPRB07,OrlitaPRL08II,OrlitaSST10,HenriksenPRL10,CrasseePRB11,OrlitaPRL11,BooshehriPRB12}.
A typical response in the quantum regime closely reflects the properties of the LL spectrum:
\begin{equation}\label{eq:LLfan}
E_i = {\rm sign}(i)v_{\rm F}\sqrt{2e\hbar|i| B},\,\, i=0, \pm1, \pm2\ldots.
\end{equation}
Inter-LL excitations follow a clear $\sqrt{B}$-dependence, and the CR absorption gains a multi-mode character~\cite{NeugebauerPRL09}
due to the non-equidistant level spacing. In the low-temperature limit we always get at least two CR absorption modes
unless the Fermi level is placed precisely between two LLs.

A closer inspection of the data in Figs.~\ref{SPKT} and~\ref{ColorPlot}(a) reveals broadening of the CR line that appears at magnetic fields
of $B_m \approx 27, 20, 16, 13.5$~T and possibly also at 11.5~T. This broadening is
best manifested as a decrease in the peak intensity, see Figs.~\ref{SPKT} or \ref{ColorPlot}, whereas the total peak area remains fairly
constant, see Fig.~\ref{Fitting}(b). This sequence of field values matches the simple rule $m B_m ={\rm const}$
(with $m=3,4,5,6$ and 7) and allows us to identify the filling
factors $\nu=-4m$, which correspond to the half-filled last occupied LL.
The broadening at fields $B_m$ thus reflects two CR modes, namely the transitions $L_{-m} \rightarrow L_{-m+1}$ and
$L_{-m-1} \rightarrow L_{-m}$,  with nearly equal intensities. It is worth noting that the CR width oscillating with the filling factor
might also be reminiscent of another phenomenon -- the difference in the screening properties of a full or half-filled last-occupied LL
in a 2D parabolic-band electron gas~\cite{AndoJPSJ75,EnglertSSC83,EnglertPhysBC83,PotemskiSS92}. Apart from the discussed
weak oscillations in $\nu$, the CR width does not exhibit any clear field-dependence, which would facilitate the identification of the main source of scattering
in the sample, based on predictions of available theories, see, e.g., Refs. \cite{ShonJPSJ98,YangPRB10,YangPRB10II}.

The sequence of the CR line broadening $m B_m ={\rm const}$ is compatible with two possible scenarios. Either the Fermi level in graphene
is pinned by electronic states in the SiC substrate, {\it i.e.}, $E_{\rm F} = {\rm const} = -v_{\rm F}\sqrt{2e\hbar m B_m}$, or the carrier density $n$ simply remains constant with varying $B$. In the latter case, the density is given by $|n| = {\rm const} = N_{\rm f} m\zeta(B)$, where $\zeta(B)$ is the LL degeneracy $eB/h$ and $N_{\rm f} =4$ is the number of fermion flavors in graphene ({\it i.e.}, a degeneracy factor associated with spin and valley degrees of freedom). Even though charge transfer between the substrate and graphene cannot be excluded ($E_{\rm F} ={\rm const}$), this effect is negligible as testified by the CR weight being nearly constant with $B$, see Fig.~\ref{Fitting}(b). Anyway, in both cases, the carrier density at $B=0$ and also at each $B_m$ is $|n|= 4 e m B_m/h = (7.9\pm 0.2)\times 10^{12}~{\rm cm}^{-2}$. Importantly, this value of carrier density is extracted entirely from the LL occupation, {\it i.e.}, independently of $v_{\rm F}$ and $E_{\rm F}$. This represents a significant advantage over previous studies of graphene in the quasi-classical regime~\cite{WitowskiPRB10,CrasseeNP11}.
We also note that the observation of individual filling factors in data, revealed as well-defined steps in position of the CR line in Fig.~\ref{ColorPlot}, serves as an indication of high homogeneity of the carrier density on the macroscopic scale.

Using the ESP expression for the cyclotron frequency, $\omega_c=eB/m_{\rm c}$, we can extract the cyclotron mass at the Fermi level from
the slope of the CR line directly: we find $m_{\rm c}=(0.058\pm0.01)m_{\rm e}$, where $m_{\rm e}$ is the electron mass in vacuum.  Using the ``Einstein relation'', $|E_{\rm F}| = m_{\rm c} v_{\rm F}^2$, together with $|E_{\rm F}| = \hbar v_{\rm F}\sqrt{\pi |n|}$, we obtain $v_{\rm F}$ and $E_{\rm F}$.
We find $v_{\rm F}=(0.99 \pm 0.02)\times 10^{6}~{\rm m}/{\rm s}$, which is sensibly larger than the ``bare" value $v_{{\rm F}, 0}$, and $|E_{\rm F}|= (325 \pm 5)~{\rm meV}$ which is
in excellent agreement with the position of onset of the interband absorption (not discussed in this paper), which provides $|E_{\rm F}|=(320\pm10)$~meV.
According to the ESP model (\ref{eq:singleparticlelike}), the deduced Fermi level implies the expected Drude weight ${\cal D}\hbar/\sigma_{\rm uni} = 2|E_{\rm F}| = 650~{\rm meV}$.

On the other hand, the Drude weight can be directly extracted from our data by estimating the area under
$\Re e~[\sigma_{xx}(\omega,B)]$, which bypasses the use of Eq.~(\ref{eq:singleparticlelike}). To account for the partial transfer of the Drude weight into
the lower magneto-plasmon branch~\cite{AllenPRB83,HeitmannSS92,HeitmannPT93}, which is significant at lower magnetic fields, the Drude weight has been extracted
by fitting with a standard Drude-Lorentz formula, see Refs. \cite{CrasseeNL12,YanNL12}, assuming the confined-plasmon frequency of $\omega_0=6.5$~meV \cite{CrasseeNL12}.
This approach provides at higher fields results identical to the fitting with a simple Lorentzian curve. From this fitting, we obtained the Drude weight close to ${\cal D}\hbar/\sigma_{\rm uni} \approx 600~{\rm meV}$, see Fig.~\ref{Fitting}(b).

To compare both estimates of the Drude weight, the following corrections should be taken into account: i) the coverage of the substrate by graphene is not full
(see, {\it e.g.}, AFM measurements on similarly-prepared samples~\cite{CrasseeNL12}), ii) the hydrogenization process is not always complete -- a small part of the sample
remains covered only by the non-graphene ``zero layer''~\cite{FortiPRB11}, iii) bilayer graphene may also appear at selected locations~\cite{Ostlerpssb10} and
iv) the area below $\Re e~[\sigma_{xx}(\omega,B)]$, extracted using the simplified Eq.~\ref{Tinv}, is suppressed with respect to the Drude weight
by a few percent, see Appendix. All these corrections tend to decrease the Drude weight deduced from the area under $\Re e~[\sigma_{xx}(\omega,B)]$ by approximately 5-10\%.
To sum up, we find that our two independent estimates of the Drude weight agree with each other with a precision better than 10\% and conclude that we do not observe any significant deviation from the validity of the ESP model (\ref{eq:singleparticlelike}). Our findings thus do not support recent transmission studies~\cite{HorngPRB11,YanACSNano11} in which a significant suppression of the Drude weight (in comparison with the ESP expectation) has been reported. The reason for this discrepancy remains unclear at the moment, although we believe that (at least a part of) the suppression of the Drude weight found in Ref.~\cite{HorngPRB11} might stem from the used normalization procedure, in which the Drude-type absorption of graphene in the regime of electron-hole puddles (when the sample is neutral on average) is neglected.

\section{Discussion}

The validity of the ESP model (\ref{eq:singleparticlelike}) for the description of the strength of intraband absorption reminds us of Landau's theory of normal Fermi liquids, which establishes a ``mapping" between a system of interacting fermions and a gas of weakly-interacting quasiparticles with renormalized parameters~\cite{Giuliani_and_Vignale}. The renormalized quasiparticle velocity $v_{\rm F}$ is precisely one of these parameters, which is completely controlled by the real part of the quasiparticle self-energy evaluated at the Fermi level $E_{\rm F}$. Quasiparticles, however, do interact among each other and their interactions, which are described by the dimensionless Landau parameters $F^{{\rm s}, {\rm a}}_\ell$, modify physical observables such as the compressibility and the spin susceptibility, or, more in general, the macroscopic response of the interacting electron liquid to external fields~\cite{Giuliani_and_Vignale}. Our experimental test of the validity of the ESP model (\ref{eq:singleparticlelike}) tells us that interactions among quasiparticles in graphene, which diagrammatically are encoded in ``vertex" or ``excitonic" corrections, are weak, at least at large carrier densities.

Our experimental findings and the approximate validity of Eq.~(\ref{eq:singleparticlelike}) can also be interpreted in terms of an enhancement of the Drude weight ${\cal D}$ with respect to its non-interacting value ${\cal D}_0$, which is found to be given by ${\cal D}/{\cal D}_0 \approx v_{\rm F}/ v_{{\rm F}, 0} \approx 1.2$. This result is in excellent agreement with the theoretical predictions by Abedinpour {\it et al.}~\cite{AbedinpourPRB11}. By taking into account screening at the Thomas-Fermi level,
the authors of Ref.~\cite{AbedinpourPRB11} predicted ${\cal D}/{\cal D}_0 \approx 1.25$ at a carrier density
$|n| = 8\times10^{12}$~cm$^{-2}$ and for a dimensionless coupling constant $\alpha_{\rm ee} \approx 0.5$ -- for definition of $\alpha_{\rm ee}$ see, \textsl{e.g.}, Ref.~\cite{CastroNetoRMP09}. Note that this is the appropriate value of $\alpha_{\rm ee}$ for our sample, as deduced from a recent ARPES study carried out by Bostwick {\it et al.}~\cite{BostwickScience10} in a nearly identical sample. We stress that the theoretical proof of the approximate validity of Eq.~(\ref{eq:singleparticlelike}) given in Ref.~\cite{AbedinpourPRB11} is far from trivial: it is based on a fully-microscopic diagrammatic approach in which substantial cancelations between the self-energy and vertex corrections play an essential role. In particular, the authors of Ref.~\cite{AbedinpourPRB11} demonstrated the smallness of excitonic corrections to the intraband response of a doped graphene sheet, which tend to suppress the Drude weight -- see Eqs.~(21) and~(23) in Ref.~\cite{AbedinpourPRB11}. Our experimental study corroborates this finding too.

For the sake of completeness, we should note that the Drude weight ${\cal D}$ is strictly fully transferred into the CR strength only in the quasi-classical limit. In the quantum regime, the total CR weight ${\cal D}_{\rm CR}$ may differ
from ${\cal D}$. When the Fermi level is placed just in between two LLs ($|\nu|=4m+2$, $m=1,2,3\ldots$), we still have ${\cal D} = {\cal D}_{\rm CR}$, but for
the half-filled last occupied LL ($|\nu|=4m$, $m=1,2,3\ldots$) we obtain ${\cal D}_{\rm CR} < {\cal D}$~\cite{GusyninPRL07}:
\begin{equation}
{\cal D}_{\rm CR} = {\cal D}\left[\frac{1}{2}+\frac{1}{4}\left(\sqrt{1+\frac{1}{m}}+\sqrt{1-\frac{1}{m}}\right)\right]~.
\end{equation}
However, one can easily verify that this suppression, which is related to the transfer of the oscillator strength between intra- and inter-band
absorptions,  remains very small for $m\geq 2$ (below 2\%). The approximation ${\cal D} = {\cal D}_{\rm CR}$ is thus very well applicable as long as the zero LL is not involved in the CR absorption.

\section{Conclusions}

To conclude, an absolute magneto-transmission experiment on highly-doped quasi-freestanding graphene allowed us to follow the oscillator strength
of the cyclotron resonance, which exactly matches the zero-field Drude weight. We found that this weight remains very close to ${\cal D} = 2|E_{\rm F}| \sigma_{\rm uni}/\hbar = e^2 v_{\rm F}\sqrt{\pi |n|}/(2\hbar) =  \pi |n| e^2/(2m_{\rm c})$ and that a single-particle-like picture is fully
sufficient to account for our data, provided that the renormalized (quasiparticle) parameters $E_{\rm F}, v_{\rm F}$ and/or $m_{\rm c}$ are considered. This implies that the Drude weight in a doped graphene sheet is enhanced with respect to its bare value due to electron-electron interactions and that the enhancement is tied to the Fermi velocity enhancement~\cite{GonzalesPRB99,EliasNaturPhys11}. Our findings are in excellent agreement with recent theoretical work~\cite{AbedinpourPRB11}, which predicted an enhancement of the Drude weight in doped graphene stemming from the interplay between broken Galilean invariance  and electron-electron interactions.

\ack

We acknowledge valuable discussions with A. H. MacDonald, G. Vignale, and Li Yang. This work has been supported by projects GACR No.~P204/10/1020, GRA/10/E006 within the ESF EuroGraphene programme (EPIGRAT) and by the Swiss National Science Foundation (SNSF), Grant 200021-120347, through the National Centre of Competence in Research ``Materials with Novel Electronic Properties-MaNEP''. We also acknowledge funding received from EuroMagNETII under the EU contract No.~228043. Work in Erlangen was supported by the DFG  and the ESF within the
EuroGraphene project GRAPHIC-RF and by the European Union within the project ConceptGraphene. M.P. was supported by the Italian Ministry of Education, University, and Research (MIUR) through the program ``FIRB - Futuro in Ricerca 2010" Grant No. RBFR10M5BT (``PLASMOGRAPH: plasmons and terahertz devices in graphene").

\appendix

\section{Extraction of the optical conductivity from transmission spectra}

\begin{figure*}
\begin{center}
    \includegraphics[width=8cm]{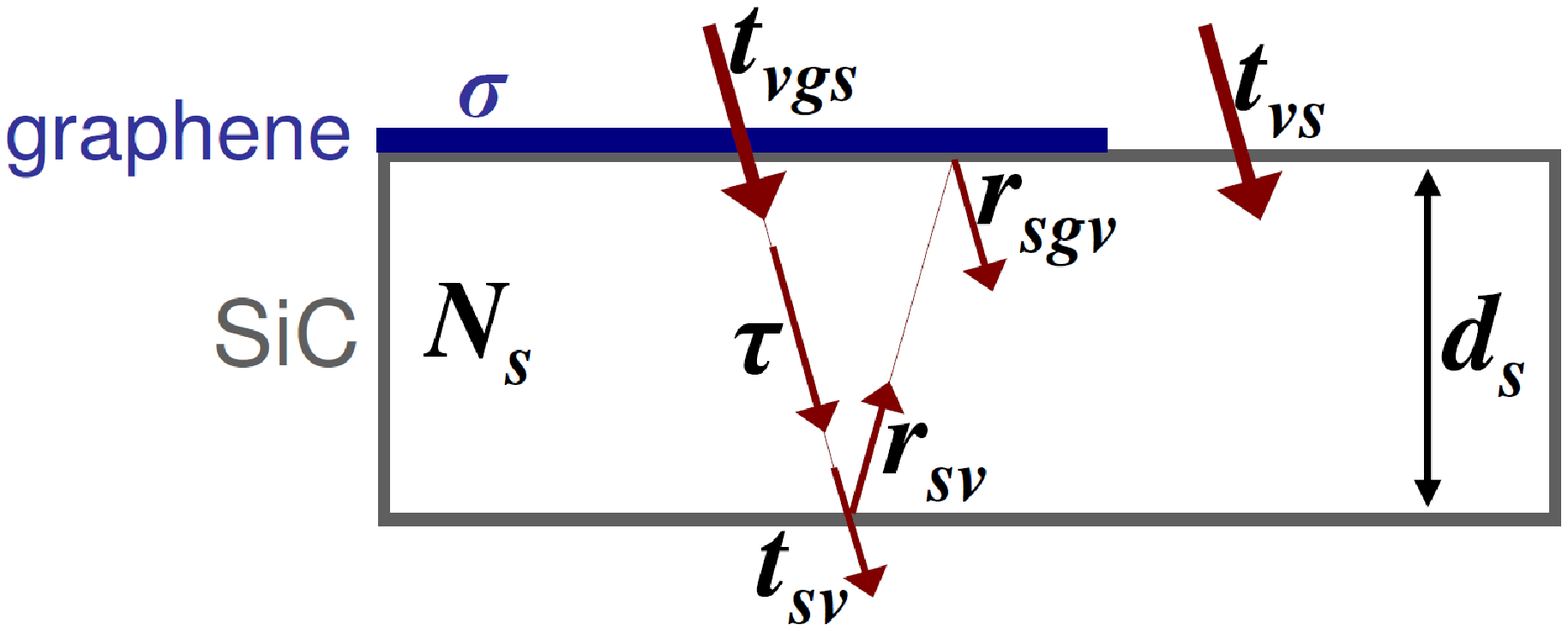}\\
    \caption{The schematic representation of the sample and the definitions of the complex coefficients of Eqn.
    (\ref{coefs}). A near normal incidence was used, the rays on the figure are inclined for clarity.}
    \label{Fig1}
\end{center}
\end{figure*}

The experimentally obtained substrate-normalized transmission $T(B)$ spectra
are used to extract the optical conductivity, $\mbox{Re}
\{\sigma_{xx}(\omega,B)\}$, of the graphene on the SiC substrate at every
field. Figure \ref{Fig1} shows the complex transmission and (internal)
reflection coefficients for our experimental ``vacuum ($v$) / graphene ($g$)
/ substrate ($s$) / vacuum ($v$)" geometry, which are:
\begin{eqnarray}
t_{vs} &=& \frac{2}{N_s + 1}, \: t_{sv} = \frac{2N_{s}}{N_s + 1}, \, t_{vgs} = \frac{2}{N_s + 1 + Z_{0}\sigma},\nonumber\\
r_{sv} &=& \frac{N_{s} - 1}{N_s + 1}, \: r_{sgv} = \frac{N_{s} - 1 - Z_{0}\sigma}{N_s + 1+ Z_{0}\sigma}, \: \tau =
\exp(i\omega N_{s}d_{s}/c).
\label{coefs}
\end{eqnarray}
\noindent Here $N_{s} = n_{s} + i k_{s}$ is the experimentally known complex
refractive index and $d_{s}$ is the thickness of the substrate, $Z_{0}$ = 377
$\Omega$ is the impedance of vacuum and $\sigma$ is the two-dimensional
optical conductivity of graphene.

In the derivation of the ratio of the transmission coefficient of the sample
to the one of the bare substrate all internal reflections in the substrate
must be taken into account. The experimental spectral resolution was reduced
to 4 cm$^{-1}$ in order to suppress the Fabry-Perot interference in the SiC
substrate. In this case, the internal reflected rays add incoherently. By
plane-wave counting, the experimental transmission coefficients of the bare
substrate and the sample are derived in terms of the complex coefficients:
\begin{eqnarray}
T_{vsv} = \frac{\left|t_{vs}t_{sv}\tau\right|^2}{1 - \left|r_{sv}^2\tau^2\right|^2}, \;
T_{vgsv} =  \frac{\left|t_{vgs}t_{sv}\tau\right|^2}{1 - \left|r_{sgv}r_{sv}\tau^2\right|^2}.
\label{transcoefs}
\end{eqnarray}
\noindent The substrate-normalized transmission follows from Eqns.
(\ref{coefs}) and (\ref{transcoefs}):
\begin{eqnarray}
T = \frac{T_{vgsv}}{T_{vsv}} = \frac{|N_{s} + 1|^4 - |N_{s} - 1|^4|\tau|^4}{|(N_{s} + 1)(N_{s} + 1 + Z_{0}\sigma)|^2 -
|(N_{s} - 1)(N_{s} - 1 -
Z_{0}\sigma)|^2|\tau|^4}.
\label{trans}
\end{eqnarray}
\noindent In the experimental spectral range, $k_{s} \ll n_{s}$, therefore
$k_{s}$ is neglected in all complex coefficients (\ref{coefs}), except in
$\tau$. We introduce the absorption coefficient of the substrate $\alpha =
|\tau|^2$ and the substrate-vacuum amplitude reflectivity $r =
(n_{s}-1)/(n_{s}+1)$ in Eqn. (\ref{trans}) and after some algebra the
relation between the substrate-normalized transmission and the complex
optical conductivity is found:
\begin{equation}
T = \left[1 + \frac{2}{n_{s} + 1}\frac{1 + \alpha^2 r^3}{1 - \alpha^2 r^4}\:\mbox{Re}\{Z_{0}\sigma\}
+ \frac{1}{(n_{s} + 1)^2}\frac{1 - \alpha^2 r^2}{1 - \alpha^2 r^4}\left|Z_{0}\sigma\right|^2\right]^{-1}.
\label{Tinv}
\end{equation}
Evidently, from only the experimental transmission spectra the complex
conductivity $\sigma$ can not be extracted. However, note that the prefactor
in the term depending on $\mbox{Re}\{\sigma\}$ is significantly larger than
the prefactor in the term depending on $|\sigma|^2$, therefore the
transmission is determined predominantly by $\mbox{Re}\{\sigma\}$. In this
case replacing $|\sigma|^2$ with $\mbox{Re}\{\sigma\}^2$ is a reasonable
approximation. In magnetic field, Eq. (\ref{Tinv}) remains valid for each
circular polarization separately, if $\sigma$ is substituted with
$\sigma_{\pm}$. For unpolarized light used in our experiment, the
transmission is an average of the two. In order to extract the diagonal
conductivity at finite field, we made a further simplification by ignoring
the relatively small effect of $\sigma_{xy}$ on the transmission and
substituting $\sigma$ with $\mbox{Re} \{\sigma_{xx}\}$ in Eqn. (\ref{Tinv}).
To estimate the error introduced by these approximations,
we have tested the simplified version of equation~\ref{Tinv} using the ideal classical
CR absorption described by the standard analytical formulae
for $\sigma_{xx}$ and $\sigma_{xy}$. We have found that the area
under $\mbox{Re} \{\sigma_{xx}\}$, extracted using the above described
procedure, is suppressed with respect to the Drude weight by roughly 5\%.

\section*{References}


\end{document}